\documentclass[twocolumn,amssymb,amsmath,floats,showpacs,superscriptaddress,prl,floatfix]{revtex4-1}

\usepackage{amsmath,amssymb,bm}
\usepackage{graphics}
\usepackage{epsfig}
\usepackage{graphicx}

\begin{document}

\title{Localization and spreading of diseases in complex networks}

\author{A. V. Goltsev}
\affiliation{
Department of Physics $\&$ I3N, University of Aveiro,
3810-193 Aveiro, Portugal}
\affiliation{Ioffe Physico-Technical Institute, 194021 St. Petersburg, Russia}
\author{S. N. Dorogovtsev}
\affiliation{
Department of Physics $\&$ I3N, University of Aveiro,
3810-193 Aveiro, Portugal}
\affiliation{Ioffe Physico-Technical Institute, 194021 St. Petersburg, Russia}
\author{J. G. Oliveira}
\affiliation{
Department of Physics $\&$ I3N, University of Aveiro,
3810-193 Aveiro, Portugal}
\affiliation{Departamento de Engenharia F\'{\i}sica, Faculdade de Engenharia,
Universidade do Porto, rua Dr. Roberto Frias, 4200-465 Porto, Portugal}
\author{J. F. F. Mendes}
\affiliation{
Department of Physics $\&$ I3N, University of Aveiro,
3810-193 Aveiro, Portugal}

\begin{abstract}
Using the SIS model on unweighted and weighted networks, we consider the disease localization phenomenon. In contrast to the well-recognized point of view that diseases infect a finite fraction of vertices right above the epidemic threshold, we show that diseases can be localized on a finite number of vertices, where hubs and edges with large weights are centers of localization.
Our results follow from the analysis of standard models of
networks and empirical data for real-world networks.

\end{abstract}

\pacs{05.10.-a, 05.40.-a, 05.50.+q, 87.18.Sn}
\maketitle





Survey of infectious diseases reveals that
before an outbreak, often, if not typically, a disease is localized within a small group of individuals.
Changes in environmental conditions or increase
in the frequency of external contacts result in an epidemic outbreak.
In the present paper we propose an approach that
enables us to describe quantitatively this important localization-delocalization phenomenon.
Our approach is based on the
SIS model \cite{ab01a,psv2001a} of spreading of diseases in weighted and unweighted
networks, where the weights of edges encode frequency of contacts between vertices.
It is widely accepted that in uncorrelated networks the epidemic threshold $\lambda_{c}$ of the infection rate $\lambda$ is $\lambda_{MF} =\langle q \rangle / \langle q^2 \rangle$,  where $\langle q \rangle$ and $\langle q^2 \rangle$ are the first and second moments of the degree distribution \cite{psv2001a}. So in networks with a finite $\langle q^2 \rangle$ the threshold should be non-zero, while it is zero if $\langle q^2 \rangle$ diverges.
One should stress however that all these well-known results were obtained only within a mean-field theory,
actually within an annealed network approximation in which a random network is substituted for its fully connected weighted counterpart \cite{psv2001a}.
Contrastingly, one can show exactly for an arbitrary graph that $\lambda_{c}$ is actually determined by the largest eigenvalue $\Lambda_1$ of the adjacency matrix $A_{ij}$ of the graph, and $\lambda_{c}=1/\Lambda_1 <\lambda_{MF}$ \cite{Wang2003,Chung2003,dgms2003,Gomez2010,Prakash2010,cps2010,cp2011,mbps2010,Kitsak2010}.
For uncorrelated networks, in particular, scale-free networks with the degree exponent $\gamma >2.5$, it was found that $\Lambda_1$ is determined by the maximum degree $q_{max}$, $\Lambda_1 \propto \sqrt{q_{max}}$ \cite{Wang2003,Chung2003,dgms2003}.
%
Then, if in the infinite size limit,
$q_{max}$ tends to infinity, as, e.g., in the Erd\H{o}s-R\'{e}nyi graphs, this leads to the amazing conclusion that the epidemic threshold is absent even in (infinite) networks
with a finite $\langle q^2 \rangle$ in contrast to the mean-field result.
The conclusion that the epidemic threshold may be absent even in the networks with rapidly decaying degree distributions was confirmed in numerical simulations performed in Ref. \cite{cps2010}.

In the present paper we
develop a spectral approach to the SIS model on complex networks. We show that the contradiction between the mean-field approximation and the exact result can be resolved if we take into account localization of diseases.
It turns out that, in contrast to the mean field theory, in which a finite fraction of vertices are infected at $\lambda >\lambda_c$, there are actually two scenarios of the spreading of diseases.  If $\Lambda_1$ corresponds to a localized eigenstate, then, at
$\lambda$ right above $\lambda_{c}=1/\Lambda_1$, disease is mainly localized on a finite number of vertices, i.e., the fraction of infected vertices is negligibly small in large networks.
With further increase of $\lambda$, the
disease gradually infects more and more vertices until
it will infect a finite fraction of vertices.
In the second scenario, $\Lambda_1$ corresponds to a delocalized state.
Then already at $\lambda \Lambda_1{-}1{\ll}1$, the disease infects a finite fraction of vertices.
Analysing network models and real-world networks,
we show that hubs, edges with large weights, and other dense subgraphs
can be centers of localization.

We consider the standard SIS model of disease spreading in a complex network of size $N$
having adjacency matrix with arbitrary entries $A_{ij}\geq 0$.
Infected vertices become susceptible with unit rate, and each susceptible vertex becomes infected by its infective neighbor with the infection rate $\lambda$.
Neglecting correlations between infected and susceptible vertices, the probability $\rho_{i}(t)$ that vertex $i$ is infected at time $t$ is
described by the evolution equation
\begin{equation}
\frac{d\rho_{i}(t)}{dt}=-\rho_{i}(t) +\lambda[1-\rho_{i}(t)]\sum_{j=1}^{N}A_{ij}\rho_{j}(t).
\label{SIS1}
\end{equation}
In the steady state,
at $t\rightarrow \infty$, the
probability $\rho_i \equiv \rho_i(\infty)$ is determined by a non-linear equation,
\begin{equation}
\rho_{i}=\frac{\lambda\sum_{j}A_{ij}\rho_{j}}{1+\lambda\sum_{j}A_{ij}\rho_{j}},
\label{SIS2}
\end{equation}
which has a non-zero solution $\rho_{i} >0$ if $\lambda$ is larger than the so-called
epidemic threshold $\lambda_{c}$. In this case,
the prevalence $\rho\equiv\sum_{i=1}^{N}\rho_{i}/N$ is non-zero.

\emph{Spectral approach.}---To solve the SIS model, we
use the spectral properties of the adjacency matrix $\widehat{A}$. The eigenvalues $\Lambda$ and the corresponding eigenvectors
$\mbox{\boldmath$f$}$
with components $f_{i}$ are solutions of the equation
$\Lambda\mbox{\boldmath$f$}=\widehat{A}\mbox{\boldmath$f$}$.
Since the matrix $\widehat{A}$ is real and symmetric,
its $N$ eigenvectors
$\mbox{\boldmath$f$}(\Lambda)$ ($\Lambda_{max} \equiv \Lambda_{1} \geq \Lambda_{2} \geq \dots \Lambda_{N}$)
form a complete orthonormal basis.
According to the Perron-Frobenius theorem, the largest eigenvalue $\Lambda_{1}$ and the corresponding principal eigenvector
$\mbox{\boldmath$f$}(\Lambda_1)$
of a real nonnegative symmetric matrix are nonnegative \cite{Minc}.
The probabilities $\rho_i$ can be written as a linear superposition,
\begin{equation}
\rho_i=\sum_{\Lambda} c(\Lambda) f_{i} (\Lambda).
\label{exp}
\end{equation}
The coefficients $c(\Lambda)$ are the projections of the vector {\boldmath$\rho$}
on $\mbox{\boldmath$f$}(\Lambda)$.
Substituting Eq.~(\ref{exp}) into Eq.~(\ref{SIS2}), we obtain
\begin{equation}
c(\Lambda)=\lambda \sum_{\Lambda'} \Lambda' c(\Lambda') \sum_{i=1}^N \frac{f_i (\Lambda) f_i (\Lambda')}{1+\lambda \sum_{\widetilde{\Lambda}} \widetilde{\Lambda} c(\widetilde{\Lambda})f_i (\widetilde{\Lambda})}.
\label{SIS4}
\end{equation}
In order to find the epidemic threshold $\lambda_c$ and
$\rho(\lambda)$ near $\lambda_c$, it is enough to take into account only the principal eigenvector
$\mbox{\boldmath$f$}(\Lambda_{1})$
in Eqs.~(\ref{exp}) and (\ref{SIS4}), i.e., $\rho_i \approx c(\Lambda_{1}) f_{i} (\Lambda_{1})$. Solving Eq.~(\ref{SIS4}) with respect to $c(\Lambda_{1})$ gives $\lambda_c {=}1/\Lambda_{1}$. At $\lambda{\geq}  \lambda_c$ in the first order in $\tau \equiv \lambda \Lambda_{1}{-}1 {\ll} 1$,  we find $\rho \approx \alpha_1 \tau$, where the coefficient $\alpha_1$ is
\begin{equation}
\alpha_1 = \sum_{i=1}^N f_i(\Lambda_1)/[N \sum_{i=1}^N f_i^3 (\Lambda_1)].
\label{alpha}
\end{equation}
This expression is
exact if there is a gap between $\Lambda_1$ and $\Lambda_2$ (see also Ref.~\cite{Van_Mieghem:m12}). Thus, at $\tau\ll 1$, $\rho$ is
determined by the principal eigenvector. The contribution of other eigenvectors are of the order of $\tau^2$.
Considering the two largest eigenvalues in Eq.~(\ref{SIS4}),
$\Lambda_{1}$ and $\Lambda_{2}$, and their eigenvectors, we obtain $\rho(\lambda) \approx \alpha_1 \tau + \alpha_2 \tau ^2$ and so on.

\begin{figure}
\begin{center}
\scalebox{0.25}{\includegraphics[angle=0]{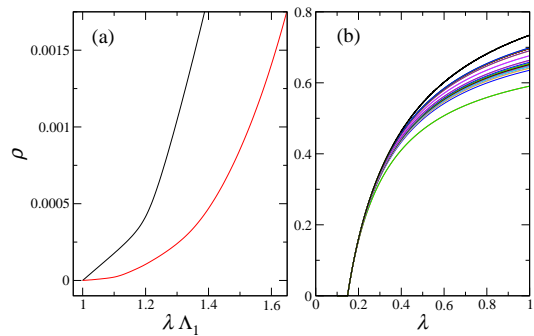}}
\end{center}
\caption{
Prevalence $\rho$ versus the infection rate $\lambda$ in real networks.
(a) astro-phys (upper line) and  cond-mat-2005 (lower line) weighted networks [from Eq.~(\ref{SIS2})]. The  eigenstate $\Lambda_1$  is localized in the cond-mat-2005 network and delocalized in the astro-phys and karate-club networks.
(b) Karate-club network. The lower curve accounts for only the eigenstate $\Lambda_1$ in Eq.~(\ref{SIS4}). Accounting for eigenstates $\Lambda_1$ and $\Lambda_2$, we find the higher curve and so on. The most upper curve is the exact $\rho$. } \label{fig-weighted}
\end{figure}

The usual point of view is that $\alpha_1$ is of the order of $O(1)$,
and so a finite fraction of vertices is infected right above $\lambda_c$.
To learn if another behavior is possible, we study whether
$\Lambda_1$ corresponds to a localized or delocalized state.
We use the inverse participation ratio
\begin{equation}
IPR(\Lambda)\equiv \sum_{i=1}^{N} f_{i}^{4}(\Lambda).
\label{IPR}
\end{equation}
If, in the limit $N \rightarrow \infty$, $IPR(\Lambda)$ is of the order of $O(1)$, then the
eigenvector {\boldmath$f$}$(\Lambda)$ is localized. If $IPR(\Lambda){}\to{} 0$ then this state is delocalized.
For a localized {\boldmath$f$}$(\Lambda)$ the components $f_{i}(\Lambda)$ are of the order of $O(1)$ only at few vertices.
For a delocalized {\boldmath$f$}$(\Lambda)$ we usually have $f_{i}(\Lambda)\sim O(1/\sqrt{N})\ll 1$. From Eq.~(\ref{alpha}) it follows that if the principal eigenvector {\boldmath$f$}$(\Lambda_1)$
is localized, then $\alpha_1 \sim O(1/N)$ and so
$\rho\approx \alpha_1 \tau \sim O(1/N)$. In this case, above $\lambda_c$
the disease is localized on a finite number $N\rho$ of vertices. If {\boldmath$f$}$(\Lambda_1)$
is delocalized, then  $\rho$ is of the order of $O(1)$,
and the disease infects a finite fraction of vertices right above $\lambda_c$. These two
contrasting scenarios
are shown in Fig.~\ref{fig-weighted} for the SIS model on the karate-club network \cite{Zachary1977} and
the weighted collaboration networks of scientists posting preprints on
the astrophysics archive at arXiv.org, 1995--1999,
and the condensed matter archive at January 1, 1995 -- March 31, 2005 \cite{Newman2001}. The astro-ph and karate-club nets have delocalized principal eigenstates while the cond-mat-2005 net has a localized principal eigenstate.
Numerical solution of Eq.~(\ref{SIS2}) gives $\alpha_1{=}1.8{\times}10^{-3}$ for the astro-ph net and smaller $\alpha_1=1.5{\times}10^{-4}$ for the cond-mat-2005 net.

One can find $\Lambda_1$ and $IPR(\Lambda_1)$ for any unweighted and weighted graph:
\begin{eqnarray}
&&
\Lambda_{1} =\lim_{n \rightarrow \infty} \Lambda_{1}(n) \equiv \lim_{n \rightarrow \infty}
(\mbox{\boldmath$g$}^{(n)} \widehat{A} \mbox{\boldmath$g$}^{(n)})/|
\mbox{\boldmath$g$}^{(n)}|^2 ,
\label{bound-2}
\\[0pt]
&&  IPR(\Lambda_1 )=\lim_{n \rightarrow \infty} \sum_{i=1}^N (g^{(n)}_i)^4 / |
\mbox{\boldmath$g$}^{(n)}|^4,
\label{bound-3}
\end{eqnarray}
where $\mbox{\boldmath$g$}^{(n+1)}{=}\widehat{A}\mbox{\boldmath$g$}^{(n)}$ and  $\mbox{\boldmath$g$}^{(0)}$ is a positive vector.
$ \Lambda_{1}(n) $ is a lower bound of $\Lambda_{1}$.
In unweighted networks, i.e., $A_{ij}=0,1$, for $\mbox{\boldmath$g$}^{(0)} {=} 1 $,
the first iteration $n=1$ gives
\begin{equation}
\Lambda_{1}(1)=\frac{1}{\langle q^2 \rangle N} \sum_{i,j} q_{i} A_{ij} q_{j} = \Lambda_{MF}
+ \frac{\langle q \rangle \sigma^2 r}{\langle q^2 \rangle},
\label{bound-4}
\end{equation}
where  $\Lambda_{MF}{\equiv} \langle q^{2} \rangle /\langle q \rangle$,  $r$ is the Pearson coefficient, and \emph{\textbf{$\sigma^2 = \langle q^3 \rangle/\langle q \rangle - \langle q^2 \rangle^2/\langle q \rangle^2$}}
\cite{Newman02,dfgm2010}.
Eq.~(\ref{bound-4})
shows that assortative degree-degree correlations ($r>0$) increase $\Lambda_{1}$ while  disassortative correlations ($r<0$) decrease $\Lambda_{1}$. The first iteration also gives the mean-field result $IPR=\langle q^4 \rangle/[N \langle q^2 \rangle ^2]\sim O(1/N)$.
A few iterations already give good approximations for $\Lambda_{1}$ and $IPR$ if the principal eigenstate is delocalized but more iterations are needed if this eigenstate is localized.

\begin{figure}[t]
\begin{center}
\scalebox{0.25}{\includegraphics[angle=0]{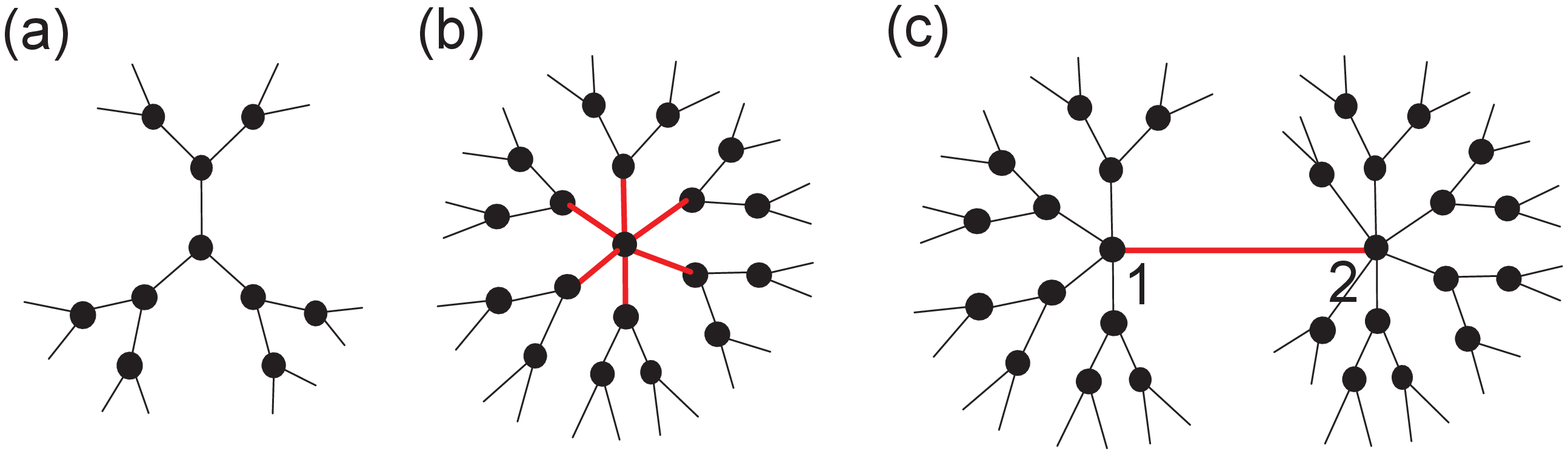}}
\end{center}
\caption{(a) Regular Bethe lattice with degree $k=3$. (b) Bethe lattice with one hub of degree $q > k$. This hub is connected to neighbors by edges having the same weight $w \geq 1$ (red lines). (c) Bethe lattice with two vertices of degrees $q_1$ and $q_2$ connected by an edge with a weight $w\geq 1$ (red line).  } \label{bethe-hubs}
\end{figure}

\emph{Bethe lattice.}---To find possible centers of localization of
$\Lambda_1$, we use
Bethe lattices as simple but representative examples of networks.
The adjacency matrix of an unweighted regular Bethe lattice in Fig.~\ref{bethe-hubs}(a) with vertices of degree $k$ has the largest eigenvalue $\Lambda_1 {=} k$ with a delocalized eigenvector $f_{i}(\Lambda_1){=}N^{-1/2}$.
Let us introduce a hub of degree $q{>}k$ connected to the neighbors by edges with a weight $w \geq 1$ [see Fig.~\ref{bethe-hubs}(b)]. The other edges have weight 1.
We look for such a solution $\mbox{\boldmath$f$}$ of the equation $\Lambda \mbox{\boldmath$f$} =\widehat{A} \mbox{\boldmath$f$} $ that has a maximum component $f_{0}(\Lambda_{1})$ at the hub and exponentially decreases with increasing distance $n$ from the hub, $f_{i} (\Lambda_{1})=f_n (\Lambda_{1})\propto 1/a^n$. We find
\begin{eqnarray}
&&
\Lambda_{1}=qw^{2}/\sqrt{qw^{2}-B},
\label{hub1}
\\[2pt]
&&
IPR(\Lambda_{1})=f_{0}^{4}(\Lambda_{1})[1 + qw^4/(a^4 - B)],
\label{hub2}
\\[2pt]
&&
f_{0}(\Lambda_{1})=[(qw^{2}/2-B)/(qw^{2}-B)]^{1/2},
\label{hub3}
\\[2pt]
&&
f_{n}(\Lambda_{1})=wf_{0}(\Lambda_{1})/a^n.
\label{hub4}
\end{eqnarray}
Here $B\equiv k-1$ is the branching coefficient of the graph, $a\equiv(qw^2-B)^{1/2}$.
Due to the exponential decay, $IPR$ is finite,
so this eigenstate is localized. In the limit $qw^{2} \gg B$, we have $IPR \rightarrow (1+1/q)/4$.
This solution gives the maximum eigenvalue if $\Lambda_{1}> k$. This condition  can be written in the form  $q > q_{loc} \equiv (B^2 +B)/w^{2}$. The second eigenstate with $\Lambda_2 = k$ and $f_{i}(\Lambda_2){\approx}N^{-1/2}$ is delocalized.

Now we consider a Bethe lattice with two hubs of degrees $q_1$ and $q_2$ connected by an edge with weight $w \geq 1$ [see Fig.~\ref{bethe-hubs}(c)]. Other edges have weight 1.
As above, we look for an eigenvector $\mbox{\boldmath$f$}$ that exponentially decays from these hubs. We find that there are two localized eigenstates with eigenvalues $\Lambda_1$ and $\Lambda_2$ above $\Lambda_3=k$,
\begin{eqnarray}
&&
\Lambda_{1(2)}=a_{\pm}+B/a_{\pm},
\nonumber
\\[2pt]
&&
a_{\pm}^2=\frac{1}{2}(Q_{1}{+}Q_{2}{+}w^{2}){\pm}\frac{1}{2}[(Q_{1}{+}Q_{2}{+}w^2)^2{-}4Q_{1}Q_{2}]^{1/2},
\nonumber
\\[2pt]
&&
\Psi_1^2(a_\pm^2{+}Q_1)+\Psi_2^2(a_\pm^2{+}Q_2) = a_{\pm}^2-B
,
\nonumber
\\[2pt]
&&
\!\!
IPR(\Lambda_{1(2)}){=}[\Psi_{1}^{4}(a_{\pm}^{4}{+}Q_1){+}\Psi_{2}^{4}(a_\pm^4{+}Q_2)]/(a_\pm^4{-}B).
\label{two-hubs1}
\end{eqnarray}
The signs $\pm$ correspond to $\Lambda_1$ and $\Lambda_2$, respectively, and $Q_{1(2)}\equiv q_{1(2)}-B-1$. The components $f_i$
decrease exponentially as $\Psi_{1(2)}/a_{\pm}^{n}$ with increasing distance $n$
from the hubs 1 and 2.
$\Psi_{1}$ and $\Psi_{2}$ are the components of $\mbox{\boldmath$f$}$ at the hubs 1 and 2. Their ratio is  $\Psi_{2}/\Psi_{1}=(a_{\pm}^{2}{-}Q_1)/(wa_{\pm})$. The criterion for localization is 
$\Lambda_{1},\Lambda_{2}>k$. If $q_1=q_2 $ and $w\gg 1$, then $\Psi_{1}=\Psi_{2}\rightarrow 1/\sqrt{2}$ and $IPR(\Lambda_{1})$ reaches the maximum value 0.5 that means localization on two hubs.
In general, $\Lambda_{1}$ can be localized in a larger cluster.

\begin{figure}
\begin{center}
\scalebox{0.25}{\includegraphics[angle=0]{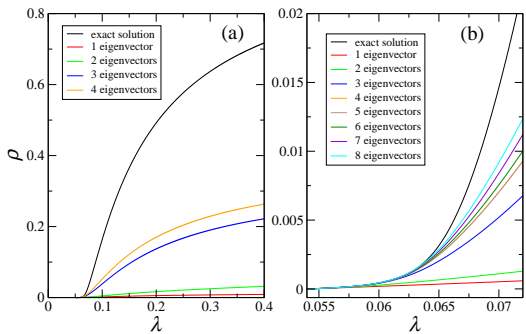}}
\end{center}
\caption{(a) Prevalence $\rho$ versus $\lambda$ in a scale-free network of
$10^5$ vertices generated by the static model with $\gamma =4$, $\langle q\rangle=10$. The lowest curve accounts for only
the principal eigenstate in Eq.~(\ref{SIS4}), the next one accounts $\Lambda_{1}$ and $\Lambda_{2}$, and so on. (b) Zoom of the prevalence at $\lambda$ near $\lambda_c=1/\Lambda_1$.}
\label{fig3}
\end{figure}

\emph{Scale-free networks.}---To study the appearance and properties of localized eigenstates in uncorrelated complex networks, we use the static model \cite{Goh:gkk01} that generates unweighted scale-free networks with degree distribution $P(q) \propto C q^{-\gamma}$ at $q\gg 1$. Using
software OCTAVE, for each realization of a random network of size $N$ with mean degree $\langle q\rangle $ and $\gamma=4$, we calculated eigenvalues, eigenvectors, and $IPR(\Lambda)$ of the adjacency matrix.
In networks of size $N=10^{5}$, we found that several (typically, from one to three for different realizations) eigenstates appear above the upper delocalized eigenstate.
These states are localized at hubs and their properties are described well by Eqs.~(\ref{hub1})--(\ref{hub4}) with $w=1$
if the branching coefficient $B$ in these equations is replaced by the averaged branching coefficient $B=\langle q^2 \rangle/\langle q \rangle {-}1$.
We observed that in these scale-free graphs, the upper delocalized eigenstate $\Lambda_{d}$ is
slightly above the mean-field value $\Lambda_{MF} = \langle q^2 \rangle / \langle q \rangle $. The maximum degree $q_{max}$ fluctuates from realization to realization. Localization of the principal eigenstate at a vertex with degree $q_{max}$ occurs if
\begin{equation}
\Lambda_{1}=q_{max}/\sqrt{q_{max}-B} \geq \Lambda_{d}.
\label{s-f-1}
\end{equation}
The equality here gives the threshold degree $q_{loc}$. In realizations with $q_{max} <q_{loc}$, the principal eigenvector is delocalized and $\Lambda_{1}=\Lambda_{d}$.
For $N=10^{5}$, $\langle q\rangle =10$, and $\gamma=4$, our numerical calculations give $\langle q^2 \rangle / \langle q \rangle \approx 14.1$ and $\Lambda_{d} \approx 15$. According to Eq.~(\ref{s-f-1}), a localized state appears above $\Lambda_{d}$ if $q_{max}$ is larger than $q_{loc} \approx 214$.
Since the average value of $q_{max}$ depends on $N$, at small $N$ the probability to generate a graph with $q_{max} > q_{loc}$ is small \cite{km07}. Only large graphs can have a localized principal eigenstate.
The criterion (\ref{s-f-1}) is not satisfied at $\gamma {\leq} 5/2$ because $\Lambda_{d}$ becomes larger than the eigenvalue $\Lambda \approx \sqrt{q_{max}}$ of a state localized at the vertex with $q_{max}$. Indeed, assuming $\Lambda_{d}\approx \Lambda_{MF}$, we find   $\Lambda_{d} \propto  q_{max}^{3{-}\gamma} > \sqrt{q_{max}}$ at $q_{max}{\gg} 1$ when $\gamma {\leq} 5/2$. Hence, the largest eigenstate is delocalized and $\Lambda_1=\Lambda_{d} \approx \Lambda_{MF}$ in agreement with Refs.~\cite{Chung2003,cps2010}.
Thus, in the case of uncorrelated random graphs of sufficiently large size, the principal eigenvector is localized if $\gamma>5/2$, which includes the Erd\H{o}s--R\'enyi graphs, and delocalized if $ 2 <\gamma \leq 5/2$.
Fig.~\ref{fig3} represents the results of our numerical solution of Eq.~(\ref{SIS4}) for the SIS model on one typical realization of the scale-free network. The principal eigenvector is localized at the hub with $q_{max}=323$. Equations (\ref{hub1})--(\ref{hub4}) and (\ref{alpha}) give $\Lambda_{1}=18.35$, $IPR=0.23$, and $\alpha_1 \simeq 1.4 {\times} 10^{-3}$. These values agree well with the measured values $\Lambda_{1}=18.47$, $IPR=0.21$, and $\alpha_1\simeq 1.7 {\times} 10^{-3}$. The eigenvector with $\Lambda_2$  is localized at the second largest hub with $q=254$. The third eigenvector with $\Lambda_3\approx 15.3$ is delocalized. The first two eigenstates allow to describe $\rho(\lambda)$ close to $\lambda_c {=}1/\Lambda_{1}$. Accounting for the delocalized eigenstate $\Lambda_3$ gives better results in a broader range of $\lambda$ (see Fig.~\ref{fig3}).

\begin{table*}[t]
\caption{
Characteristics of real-world networks. $N$ is size, $\gamma$ is the degree
distribution exponent, $q_{max}$ is the maximum degree,
$q_{loc}$ is the localization threshold found from Eq.~(\ref{s-f-1}),
$\Lambda_{1}$ is the largest eigenvalue and $\Lambda_{1}(1)$ is its lower bound,  Eq.~(\ref{bound-4}), respectively.
D and A stand for assortative and disassortative mixing. Two last columns represent weighted networks.}
    \begin{tabular}{cccccccccccc
}
\hline
&&&
\\[-11pt]
$\text{Network }$ & $N$ & $\gamma$ & $q_{max}$ & $q_{loc}$ & $\langle q^2 \rangle / \langle q \rangle$ & \text{ mixing} & $\Lambda_1$  & $\Lambda_{1}(1)$ &  $IPR(\Lambda_1)$ & $\Lambda_{1}$ &  $IPR(\Lambda_1)$
\\[-3pt]
& & & & & & & &  &   & \multicolumn{2}{c}{weighted\ \ }
\\[1pt]
\hline
&&&&
\\[-10pt]
\text{cond-mat 2005 \cite{Newman2001}} & 40421 & 3.0 & 278 & 2604 & 27.35  & A & 51.29 & 35.205  & 0.0081 & 47.63 &  0.3415
\\
\text{hep-th \cite{Newman2001}} & 8361 & $-$ & 50 & 521 & 8.687  & A  & 23 & 10.632 & 0.0417  & 40.52  & 0.3531
\\
\text{astro-ph \cite{Newman2001}}  & 16706 & $-$ & 360 & 5415 & 44.92  & A  & 73.89 & 56.287  & 0.005 &  33.7575 &  0.0525
\\
\text{power grid} \cite{ws1998} & 4941 & \text{exponential} & 19  & 53 & 3.87  &  $-$ & 7.483 & 3.9 & 0.041 & &
\\
\text{fp5} \cite{Almendral2007} & 27985 & 2.2 & 2942 & 38610 & 211.0 & $-$  & 197.03 & 176.3 & 0.0035 & &
\\
\text{CAIDA}\,\text{(router-internet)}\,\cite{CAIDA} & 192244 & 2.7 & 1071 & 11947 & 37.89  & $-$ & 109.5 & 42.9 & 0.010 & &
\\
\text{karate club} \cite{Zachary1977} & 34 & $-$ & 17 & 37 & 7.77 & D & 6.72  & 6.01  & 0.073 & &
\\[2pt]
\hline
\end{tabular}
\label{table3}
\end{table*}

\emph{Real networks.}---The largest eigenvalue $\Lambda_1$, $IPR(\Lambda_1)$, and other parameters of a few weighted and unweighted real-world networks are given in Table~\ref{table3}.
Note first that in all of these unweighted real networks the inverse participation ratio $IPR(\Lambda_1)$ is small that evidences a delocalized $\Lambda_1$.
We suggest that localization does not occur because the localization threshold $q_{loc}$  from the criterion Eq.~(\ref{s-f-1}) exceeds $q_{max}$.
Second, in unweighted networks,
$\Lambda_{1}$ differs strongly from the mean-field value $\Lambda_{MF} = \langle q^2 \rangle / \langle q \rangle $. $\Lambda_{1}$ is larger than $\Lambda_{MF}$ in networks with assortative mixing (cond-mat 2005, hep-th, and astro-ph networks) while $\Lambda_{1}$ is smaller than $\Lambda_{MF}$ in disassortative networks (karate club network). Qualitatively, this agrees with Eq. (\ref{bound-4}).
A similar observation was made in Refs.~\cite{cp2011,Kitsak2010}.
%
%
Table~\ref{table3} shows that in contrast to the unweighted hep-th and cond-mat-2005 networks, their weighted versions have a localized principal eigenvector with a large $IPR$. Localization occurs at vertices linked by edges with a large weight. In the cond-mat-2005 network, localization occurs at two vertices of degrees 37 and 28 connected by an edge with weight 34.3 that is
much larger than the average weight $\overline{w}=0.51$. In this case, Eq.~(\ref{two-hubs1}) gives $\Lambda_1 \approx 34.5$ and $IPR \approx 0.49$.
In the hep-th network, the strong edge has weight 34 larger than $\overline{w}=0.97$ and connects two vertices of degrees 34 and 33. Using Eq.~(\ref{two-hubs1}), we find $\Lambda_1 \approx 35$ and $IPR \approx 0.47$ in agreement with the data in Table~\ref{table3}. The components of the principal eigenvectors in these networks decay exponentially with distance from the strong edges in agreement with Eq.~(\ref{two-hubs1}).
In the astro-ph weighted network none of the edges satisfies the localization criterion.
Two scenarios of behavior of the prevalence $\rho(\lambda)$ in weighted networks with localized and delocalized $\Lambda_1$ are shown in Fig.~\ref{fig-weighted}(a).
Although above we considered only localization centers with one or two vertices, note that a disease may also be localized in larger finite clusters.

It was concluded in Refs.~\cite{Kitsak2010,cp2011} that in unweighted networks a disease first survives inside the higher $k$-cores. By definition, $k$-cores are subgraphs containing a finite fraction of a network, and so these two works actually discussed the delocalized state of disease. The principal difference of the present work from Refs.~\cite{Kitsak2010,cp2011} is that we consider situations in which a disease takes in a finite number of vertices and not a finite fraction both in unweighted and weighted networks.

In conclusion, based on a spectral approach to the SIS model,
we showed that if the principal eigenvector of the adjacency matrix of a network is localized, then at the infection rate $\lambda$ right above the threshold $1/\Lambda_1$, the disease is mainly localized on a finite number of vertices.
Importantly, a strict epidemic threshold in this case is actually absent,
and a real epidemic affecting a finite fraction of vertices occurs
after a smooth crossover, at higher values of $\lambda$.
On the other hand, if the principal eigenvector is delocalized, the epidemic occurs in the whole region above $\lambda_c=1/\Lambda_1$.
We suggest that further investigations of
real-world networks will give many
new examples of disease localization-delocalization phenomena.
\begin{acknowledgments}
This work was partially supported by the FCT
projects PTDC:
FIS/71551/2006, FIS/108476/2008, SAU-NEU/103904/2008, MAT/114515/2009, and PEst-C/CTM/LA0025/2011\vspace{-5pt}.
\end{acknowledgments}

\end{document}